\documentclass[article,nojss]{jss}

\usepackage[utf8]{inputenc}
\usepackage[normalem]{ulem}
\usepackage{amsmath,amsfonts,amssymb,graphicx,multirow}

\newcommand{\argmax}{\mathop{\mbox{argmax}}}

\usepackage{mdsymbol}

\author{Emmanuel Jordy Menvouta\\KU Leuven  \And 
	Sven Serneels\\Aspen Technology \And Tim Verdonck\\University of Antwerp\\KU Leuven}
\title{\pkg{direpack}: A \proglang{Python 3} package for state-of-the-art statistical dimension reduction methods}

\Plainauthor{Emmanuel Jordy Menvouta, Sven Serneels, Tim Verdonck} %% comma-separated
\Plaintitle{direpack: A Python 3 package for state-of-the-art statistical dimension reduction methods} %% without formatting
\Shorttitle{\pkg{direpack}: statistical dimension reduction methods in \proglang{Python 3}}%Python 3 dimension reduction} %% a short title (if necessary)

\Abstract{The \pkg{direpack} package aims to establish a set of modern statistical dimension reduction techniques into the \proglang{Python} universe as a single, consistent package. The dimension reduction methods included resort into three categories: projection pursuit based dimension reduction, sufficient dimension reduction, and robust M estimators for dimension reduction. As a corollary, regularized regression estimators based on these reduced dimension spaces are provided as well, ranging from classical principal component regression up to sparse partial robust M regression. The package also contains a set of classical and robust pre-processing utilities, including generalized spatial signs,  as well as dedicated plotting functionality and cross-validation utilities. Finally, \pkg{direpack} has been written consistent with the \pkg{scikit-learn} API, such that the estimators can flawlessly be included into (statistical and/or machine) learning pipelines in that framework.
}
\Plainkeywords{dimension reduction, projection pursuit, sufficient dimension reduction, robust statistics, energy statistics, statistical learning, multivariate statistics, Python 3} %% without formatting
\Address{
	Emmanuel Jordy Menvouta Npkw\'{e}l\'{e}\\
	KU Leuven\\
	Department of Mathematics, Faculty of Science\\
	Celest\ij nenlaan 200B, 3001 Leuven, Belgium\\
	E-mail: \email{emmanueljordy.menvoutankpwele@kuleuven.be}\\
	\linebreak
	Sven Serneels\\
	Aspen Technology\\
	Bedford, Massachusetts,\\ 
	MA01730, USA\\
	E-mail: \email{svenserneels@gmail.com}\\
	\linebreak
	Tim Verdonck\\
	University of Antwerp\\
	Department of Mathematics, Faculty of Science\\
	Middelheimlaan 1, 2020 Antwerp, Belgium\\
	\&\\
	KU Leuven\\
	Department of Mathematics, Faculty of Science\\
	Celest\ij nenlaan 200B, 3001 Leuven, Belgium\\
	E-mail: \email{tim.verdonck@uantwerp.be}\\
}
\begin{document}
	
	%% include your article here, just as usual
	%% Note that you should use the \pkg{}, \proglang{} and \code{} commands.
	
	\section{Introduction}\label{sec:Intro}
	According to several trackers, e.g. the popularity of programming languages index (PYPL, \url{http://pypl.github.io/PYPL.html}) \proglang{Python} is the most popular programming language. While this is not exact science, it is indicative of how widespread \proglang{Python} is adopted as a language. That statement holds for \proglang{Python} as a general purpose programming language and it is even more pronounced when focusing on data science or artificial intelligence. In spite of \proglang{Python}'s widespread adoption in the machine learning and data science communities, particularly the corporate ones, it is fair to say that the amount of particularly multivariate statistical learning methods available as open source packages in \proglang{Python}, is considerably lower than in \proglang{R}. The present publication introduces the \pkg{direpack} package, aiming to narrow that gap. The \pkg{direpack} package contains a comprehensive set of multivariate statistical dimension reduction methods as a \proglang{Python 3} package. Moreover, the package has been consistently programmed, adherent to the \pkg{scikit-learn} API, such that the estimators contained in it can flawlessly be included in \pkg{scikit-learn} machine learning pipelines, or be subjected to \pkg{scikit-learn}'s built-in libraries for hyperparameter tuning. 
	
	Dimension reduction is a key building block in statistical data analysis. It has become increasingly important over the last few decades and can be projected to keep doing so as computing power and data storage capacities increase. It has become prevalent in branches of applied statistics that generate big data, such as finance, bioinformatics and chemometrics, just to name a few. But dimension reduction is by no means limited to big data: it can be a key tool to help interpreting data that have any type of dimensionality that is hard to visualize and is therefore equally valid when the scope is to reduce dimensionality from e.g. five down to two.   
	
	While \pkg{scikit-learn} contains some well-established classical statistical dimension reduction techniques, e.g. principal component analysis (PCA) or partial least squares (PLS), it does not contain options for some of the more recently developed statistical tools, such as dimension reduction techniques based on robust or energy statistics. This is where \pkg{direpack} complements \pkg{scikit-learn}, by delivering a select, yet extensive, set of state-of-the-art statistical dimension reduction and regression techniques consitent with the latter's API. 
	
	The dimension reduction methods contained in \pkg{direpack} presently resort to three categories. At first, \pkg{direpack} contains a subpackage for projection pursuit (PP) dimension reduction, \pkg{ppdire}, which allows to elegantly switch between different estimators by simply exchanging the projection index. The second subpackage in \pkg{direpack}, \pkg{sudire}, focuses on sufficient dimension reduction (SDR), which has been a major focus area for statistical research over the last thirty years. Essentially, SDR is dimension reduction targeting to find a subspace in the predictor block that is sufficient to explain a predictand. Several approaches exist to SDR, and a few older, meanwhile mainstay approaches are accessible through \pkg{sudire}, but also some very recent methods that perform SDR based on energy or ball statistics. Up to our knowledge, for some of these more recent variants, \pkg{direpack} provides the only publicly available open source version. Thirdly, \pkg{direpack} offers the \pkg{sprm} subpackage, that brings a set of robust M estimators for dimension reduction and regression, as well as an efficient implementation of univariate sparse PLS. Moreover, \pkg{direpack} contains a set of functions for classical and robust data pre-processing, including recent developments such as generalized spatial signs, as well as ancillary functions to calculate bivariate measures of covariance and association and energy statistics. Finally, \pkg{direpack} offers a set of plot functions specific to the methods provided, as well as crossvalidation utilities compatible with \pkg{scikit-learn}'s hyperparameter tuning. 
	
	The article is organized as follows: each of the following sections will focus on one specific subpackage: \pkg{preprocessing}, \pkg{ppdire}, \pkg{sudire}, \pkg{sprm} and \pkg{crossvalidation}.  The final section will review \pkg{direpack} in the landscape of existing packages and will provide an outlook for further development. 
	
	The discourse in the article will be concise, aiming to convey the concept that defines each class of methods, and highlighting which methods resorting under it, have been implemented. The article deliberately refrains from including code snippets or extensive examples, since those are more directly accessible through \pkg{direpack}'s set of \proglang{jupyter} example notebooks. A brief overview of the available example notebooks is given in Appendix \ref{sec:app}.
	
	\section{Pre-processing}\label{sec:prepro}
	
	The first step in most meaningful data analytics projects will be to pre-process the data. 
	\subsection{Data standardization}\label{sec:standard}
	A first, well accepted way to pre-process data is to center them and scale them to unit variance on a column wise basis. This corresponds to transforming a $\mathbf{x}$ variable into z-scores:
	\begin{equation}\label{eq:standard}
	\mathbf{z} = \frac{\mathbf{x} - \hat{\boldsymbol{\mu}}}{\hat{\boldsymbol{\sigma}}}, 
	\end{equation}
	where $\hat{\boldsymbol{\mu}}$ and $\hat{\boldsymbol{\sigma}}$ are estimates of location and scale, respectively. 
	For normally distributed data, the appropriate way to accomplish this is by centering about the mean and dividing by the column wise standard deviation. This is good practice without any modification in many applications and as such, is implemented in \pkg{scikit-learn}'s \texttt{StandardScaler}. However, when the marginal distributions in the data significantly deviate from the normal, outliers could throw the result of that data standardization off, and robust or nonparametric alternatives become a more reliable choice. The literature on univariate robust statistics is extensive and has been documented in several textbooks, cf. \cite{Hampel} or \cite{MaronnaMY06}. Essentially, all robust statistics are subject to a trade-off between {\em efficiency} and {\em robustness}, which means that the variance of the estimates will increase as the estimator can resist a higher fraction of outliers. It is beyond question that the latter will happen. It just depends on how the estimator has been designed by how much. While \pkg{scikit-learn} provides highly robust nonparametric standardization in its \texttt{RobustScaler}, the estimators included therein are known to have a low statistical efficiency (these are the median for location and the interquartile range for scale). It is not the purpose of \pkg{direpack} to build an extensive library of univariate estimators of location and scale, yet since autoscaling the data is often an essential step, a few location and scale estimators have been implemented. For location, with increasing performance in terms of the robustness---efficiency trade-off, these are: the column wise median, the spatial median (also called $L_1$-median, although it minimizes an $L_2$ norm) and the $k$ step least trimmed squares (LTS, \cite{RousseeuwLeroy}) estimator. For scale, the consistency corrected median absolute deviation (MAD) and the $\tau$ estimator of scale \citep{maronna2002robust} have been included. Generally, it holds true that the more statistically efficient the estimator in these lists is, the higher its computational cost. For details on these estimators, the reader is referred to one of the aforementioned reference works. In \pkg{preprocessing}, these estimators can be accessed through its \texttt{VersatileScaler} class, which takes the names of these estimators as strings, but it will also accept functions of location and scale estimators, should the user prefer to apply other ones. \texttt{VersatileScaler} also follows the \pkg{scikit-learn} API: it contains methods to \texttt{fit}, \texttt{transform} and \texttt{fit\_transform}, as well as \texttt{inverse\_transform}. 
	
	\subsection{Spatial sign pre-processing}\label{sec:sspp}% Spatial sign / Normalization
	Besides standardizing data, it can be beneficial to transform data to some sort of signs. The {\em generalized spatial sign transformation} consists of transforming a variable $\mathbf{x}$ into
	\begin{equation}\label{eq:ss}
	\mathbf{u} = \left(\mathbf{x} - \hat{\boldsymbol{\mu}}\right) \times f\left(\mathbf{x} - \hat{\boldsymbol{\mu}}\right), 
	\end{equation}
	where the spatial sign is obtained by setting $f(x) = {\parallel x \parallel}^{-1}$ and $\parallel \cdot \parallel$ denotes the norm (in all published literature in this context, the $L_2$ norm). Since spatial sign pre-processing (SS-PP) consists of dividing the data by their Euclidean norm, it is also known as {\em normalizing} and as such, is available in \pkg{scikit-learn}'s \texttt{Normalizer}. Spatial sign pre-processing has been shown to convey moderate robustness to multivariate estimators that are entirely based on co-variance estimates, such as PCA or PLS \citep{SDV2006}. Moderate robustness means in this case that the resulting estimator can resist up to 50\% of outliers, but will have a sizeable bias even for small fractions of contamination. The reason why this happens is that the spatial sign transform projects {\em all} cases onto the unit sphere indiscriminately, which can drastically change data topology, and thereby introduce bias. To remedy this, recently, the {\em generalized} spatial sign transform has been proposed \citep{RaymaekersRousseeuw2019}. These authors examine a set of different functions $f$ that can be plugged into \eqref{eq:ss}, some of which will only transform those cases in the data that exceed a certain eccentricity threshold. These functions are the quadratic radial, ball, shell, Winsor and linear redescending (LR) functions, all of which can be accessed through \pkg{direpack}'s \texttt{GenSpatialSignPreprocessor}. \cite{RaymaekersRousseeuw2019} also show that generalized spatial sign pre-processing using the right plugin function can lead to generally good robustness properties for the estimator based on the transformed data, which in their study is limited to the covariance matrix. It is viable to assume that this result further generalizes to estimators that entirely derive from covariance estimation, such as PCA and PLS, yet this still has to be proven in this setting. 
	
	\section{Projection pursuit dimension reduction}\label{sec:ppdire}
	
	Beyond discussion, the class of dimension reduction with the longest standing history accessible through \pkg{direpack}, is projection pursuit (PP) dimension reduction. The original idea behind PP goes back to \cite{Kruskal1969}, and a good overview of PP as a methodological framework can be found in \cite{Huber1985}. Notably, the latter overview is much broader than dimension reduction; in what follows, the discussion will be limited to PP dimension reduction. 
	
	\subsection{Definition}\label{sec:pp_def}
	Let $\mathbf{X}$ be a data matrix that is a sample of $n$ cases of a $p$ variate random variable and $\mathbf{y}$ be a sample of a corresponding depending variable, when applicable. The set of projection pursuit scores $\mathbf{t}_i$ that span the columns of $\mathbf{T}$ are defined as linear combinations of the original variables: $\mathbf{T} = \mathbf{X}\mathbf{W}$, where the $\mathbf{w}_i$ are given by:
	\begin{subequations}\label{eq:ppdef_empir}
		\begin{equation}\label{eq:ppdef_ctricy}
		\mathbf{w}_i = \argmax_{\mathbf{a}} \mathfrak{P}\left(\mathbb{S}\left(\mathbf{a}^T\mathbf{X}\right)\right), 
		\end{equation}
		subject to:
		\begin{equation}\label{eq:ppdef_constrcy}
		\mathbf{w}_i^T\mathbf{X}^T\mathbf{X}\mathbf{w}_j = 0 \mbox{ and } \parallel \mathbf{w}_i\parallel_2 = 1, 
		\end{equation}
	\end{subequations}
	where $i,j \in [1,\min(n,p)]$, $j > i$  and the set $\mathbb{S} = \{\mathbf{X},\mathbf{y}\}$ if data for a dependent variable $Y$ exist and is a singleton containing $\mathbf{X}$ otherwise.
	Maximization criterion \eqref{eq:ppdef_ctricy} is very flexible and the properties of the dimension reduction accomplished according to it can vary widely, mainly dependent on the presence or absence of dependent variable data, as well as on $\mathfrak{P}$, which in the PP literature is referred to as the {\em projection index}.  
	
	\subsection{Projection Indices}\label{sec:ppidx}
	In projection pursuit, the projection index determines which method is being calculated. In \pkg{direpack}, projection pursuit can be called through the \pkg{ppdire} subpackge and class object, which allows the user to pass any function of appropriate dimensionality as a projection index. However, a set of popular projection indices deriving from (co-)moments, are provided as well through the \pkg{dicomo} subpackage. For several of these, plugging them in leads to well-established methods. They comprise: 
	\begin{itemize}
		\item Moment statistics: variance (PCA), higher order moments 
		\item Co-moment statistics: covariance (PLS), higher order co-moments 
		\item Standardized moments: skewness (ICA), kurtosis (ICA)
		\item Standardized co-moments: correlation coefficient (CCA), co-skewness, co-kurtosis
		\item Linear combinations of (standardized co-) moments. Here, the \texttt{capi.py} file in the \pkg{ppdire} subpackage delivers to co-moment analysis projection index \citep{Serneels2019}. 
		\item Products of (co-)moments. Particularly the continuum association measure has been provided, which is given by $\mathop{\mbox{cont}}(\mathbf{X},\mathbf{y}) = \mathop{\mbox{cov}}(\mathbf{X},\mathbf{y})\mathop{\mbox{var}}(\mathbf{X})^{\alpha-1}$. Using this continuum measure produces continuum regression (CR, \cite{StoneBrooks1990}). CR is equivalent to PLS for $\alpha = 1$ and approaches PCA as $\alpha \rightarrow\infty$.   
	\end{itemize}
	The \pkg{dicomo} subpackage will allow as well to compute trimmed versions of these same projection indices, leading to robust versions of these dimension reduction estimators, e.g. applying a trimming fraction in $(0,.5)$ to the continuum association projection index will produce robust continuum regression (RCR, \cite{SFCV2005}). Note also that technically, some of the estimators listed above would need to be squared in order to be used as a projection index, e.g. covariance. This is implemented accordingly in \pkg{direpack}.   
	
	\subsection{Projection pursuit optimizers}\label{sec:ppoptim}
	In the early days of PP, the idea behind it was the ability to scan all directions $\mathbf{a}$ maximizing Criterion \eqref{eq:ppdef_empir}. This essentially corresponds to a brute force optimization technique, which can be computationally very demanding. It is no coincidence that the most popular of the methods accessible through projection pursuit, are typically not being computed in a PP framework. For instance, for both PCA and PLS, Criterion \eqref{eq:ppdef_empir} can be solved analytically, leading to efficient algorithms that do not directly optimize \eqref{eq:ppdef_empir}. Moreover, mathematical optimization has come quite a way since the early days of PP. Whenever the projection index plugged in, leads to a convex optimization problem, it is advisable to apply an efficient numerical optimization technique. For that purpose, \pkg{ppdire} has the option to use \pkg{scipy.optimize}'s sequential least squares quadratic programming optimization (SLSQP). However, for projection indices based on ordering or ranking data, such as medians or trimmed (co-)moments, the problem is no longer convex and cannot be solved through SLSQP. For those purposes, the {\em grid} algorithm is included, which was originally developed to compute RCR \citep{FSCV2006}. 
	
	\subsection{Regularized regression}\label{sec:ppreg}
	While the main focus of \pkg{direpack} is dimension reduction, all dimension reduction techniques offer a bridge to regularized regression. This can be achieved by regressing the dependent variable onto the estimated dimension reduced space. The latter provides regularization of the covariance matrix, due to the constraints in \eqref{eq:ppdef_constrcy}, and allow to perform regression for an undersampled $\mathbf{X}$. The classical estimate is to predict $\mathbf{y}$ through least squares regression: 
	\begin{equation}\label{eq:regDR}
	\hat{\mathbf{y}} = \hat{\mathbf{T}}\hat{\mathbf{T}}^T\mathbf{y}, 
	\end{equation}
	which again leads to well-established methods such as principal component regression (PCR), PLS regression, etc. However, different regression methods can be applied as appropriate: \pkg{ppdire} accommodates for quantile regression and robust M regression (RM) as well, which is technically sourced from the \pkg{sprm} subpackage, described in Section \ref{sec:rm}. For a reference work on robust regression, see \cite{RousseeuwLeroy}. RM should be used in combination with robust projection indices.

	\section{Sufficient dimension reduction}\label{sec:sudire}
	
	A more recently developed take at dimension reduction is {\em sufficient} dimension reduction (SDR). Similar to PP, SDR tasks to identify a space of reduced dimension consisting linear combination of the original variables $\mathbf{T} = \mathbf{X}\mathbf{W}$. However, the SDR paradigm is to identify that space in such a way that it contains all information relevant to the dependent variable: 
	\begin{equation}
	\label{eq:SDR}
	\mathbf{y} \upvDash \mathbf{X}\  | \ \mathbf{T}.
	\end{equation}
	The space satisfying \eqref{eq:SDR} is called the {\em central subspace}. In contrast to PP, SDR is not defined in case there is no dependent variable. 
	
	\subsection{Algorithms for SDR}\label{sec:SDRalgo}
	The million dollar question in SDR is how to estimate the central subspace efficiently. A lot of research has been done over the last thirty years investigating different approaches in terms of asymptotics and assumptions made in each of the approaches. A good textbook providing an overview of approaches to SDR is \cite{LiSDR}. The subpackage \pkg{sudire} contains implementations of a broad set of these approaches. In what follows, only approaches that are provided in the package, will be discussed. 
	
	Generally speaking, SDR techniques roughly resort in three categories. At first, there is a successful set of approaches to SDR based on slicing the original space. Examples of these are sliced inverse regression (SIR, \cite{LiSIR1991}) and sliced-average variance estimation (SAVE, \cite{CookSAVE2000}).  A second group of developments has involved selective focus on certain directions, which has resulted in, among others, directional regression (DR, \cite{LiDR2007}), principal Hessian directions (PHD, \cite{LiPHD1992}) and the iterative Hessian transformations (IHT, \cite{CookIHT2002}).
	
	While all of the aforementioned methods are included in \pkg{sudire} and would merit a broader discussion, at this point we would like to highlight that \pkg{sudire} contains implementations of a more recent approach as well. The latter has, so far, resulted in three methods, all three of which share the following advantages: they do not require conditions of linearity or constant covariance, nor do they need distributional assumptions, yet they may be computationally more demanding. Slightly reminiscent of PP, this third group of SDR algorithms estimates the components as:
	\begin{subequations}\label{eq:sdrass_empir}
		\begin{equation}\label{eq:sdrass_ctricy}
		\mathbf{W}_h = \argmax_{\mathbf{B}} \mathfrak{V}^2\left(\mathbf{X}\mathbf{B},\mathbf{y}\right), 
		\end{equation}
		subject to:
		\begin{equation}\label{eq:sdrass_constrcy}
		\mathbf{B}^T\mathbf{X}^T\mathbf{X}\mathbf{B} = \mathbf{I}_h, 
		\end{equation}
	\end{subequations}
	where $\mathbf{B}$ is an arbitrary $p \times h$ matrix, $h \in [1,\min(n,p)]$.
	The main difference between SDR's criterion \eqref{eq:sdrass_empir} and PP's criterion \eqref{eq:ppdef_empir} is that in SDR, components are defined based on a {\em joint} optimization that deliveres an $h$ dimensional subspace in a single step, whereas in PP, the subspace is determined by extracting univariate components {\em sequentially}.
	
	Not unlike PP, the properties of the resulting dimension reduction method are derived from $\mathfrak{V}$. Good choices for $\mathfrak{V}$ are multivariate measures of association. All three published methods are based on these, resulting either from energy or ball statistics, and all are available in the \pkg{sudire} subpackage. The options for $\mathfrak{V}$ are: 
	\begin{itemize}
		\item distance covariance \citep{SzekelyRizzoBakirov2007}, leading to option \texttt{dcov-sdr} \citep{ShengYinDCOVSDR2016}; 
		\item martingale difference divergence \citep{ShaoYangMDD2014}, leading to option \texttt{mdd-sdr} \citep{ZhangMDDSDR2019};  
		\item ball covariance \citep{PanWXZ2019}, leading to option \texttt{bcov-sdr} \citep{ZhangChenBCOVSDR2019}, 
	\end{itemize}
	the last of which has also been reported to be statistically robust. However, in \pkg{sudire}, the \texttt{bcov-sdr} option has not been implemented as a fully functional option, but rather has been provided as code, since it depends on the \pkg{Ball} package for ball covariance and its current version appears to be difficult to install in some architectures. 
	
	\subsection{Optimization for SDR}\label{sec:SDRoptim}
	Just like for PP, numerically maximizing \eqref{eq:sdrass_empir} is a challenging optimization problem. Typically, the approach taken is to use one of the less computationally involved SDR methods as a warm start and then apply nonlinear programming to find the optimum. Prior to publication of \pkg{direpack}, each of the above three methods were only available in specific environments, without a common API. The publicly available \proglang{MATLAB} implementation of DCOV---SDR will start by computing the SIR, SAVE and DR solutions. It will use the solution from this set that yields the highest value for Criterion \eqref{eq:sdrass_ctricy} and use that estimate as a warm start for optimization through \proglang{MATLAB}'s SQP. Likewise, \citep{ZhangChenBCOVSDR2019} mention that the \proglang{R} implementation for BCOV---SDR starts from DCOV---SDR and then optimizes through an \proglang{R} wrapper around the \pkg{donlp2} optimizer. Neither of these options are available in \proglang{Python}, but the approach implemented in \pkg{sudire} is very similar: the same warm start is used as in the corresponding publications, and as a nonlinear optimizer, the \pkg{Python} binders for \pkg{IPOPT} are used. Interior point optimization (IPOPT) is a very efficient nonlinear optimization program that has gained popularity in the operations research community, and beyond \citep{WaechterBiegler2006}. Simulation results based on this implementation are on par with the ones reported in the above SDR articles. 
	
	\section{Sparse and Robust M estimators}\label{sec:sprm}
	
	The third dimension reduction subpackage, \pkg{sprm}, culminates in sparse and robust dimension reduction in the form of sparse partial robust M regression (SPRM). SPRM is a sparse and robust alternative to PLS that can be calculated efficiently \citep{SPRM}. The subpackage is organized slightly differently from the other two main subpackages. Because  SPRM combines the virtues of robust regression with sparse dimension reduction, besides the SPRM estimators itself, each of these building blocks are provided themselves as class objects that can be deployed in \pkg{sklearn} pipelines. The class objects \texttt{rm}, \texttt{snipls} and \texttt{sprm} are sourced by default when importing \pkg{direpack}.
	
	\subsection{Robust M regression}\label{sec:rm}
	M regression is a generalization of least squares regression in the sense that it minimizes a more general objective that allows to tune the estimator's robustness. In M regression, the vector of regression coefficients is defined as: 
	\begin{equation}\label{eq:critM}
	\hat{\boldsymbol{\beta}} = \mathop{\mbox{argmin}}_{\boldsymbol{\beta}}\sum_i
	\rho\left(\frac{r_i(\boldsymbol{\beta})}{\hat{\sigma}}\right) ,
	\end{equation}
	where $r_i$ are the casewise regression residuals and $\hat{\sigma}$ is a robust scale estimator thereof. The $\rho$ function defines the properties of the estimator. Identity to the least squares estimator is obtained if $\rho(r) = r^2$, but robustness can be introduced by taking a different function, for instance a function that is approximately quadratic for small (absolute) $r$, but increases more slowly than $r^2$ for larger values of $r$. Objective \eqref{eq:critM} can be solved numerically, but it is well known that that solution can equivalently be obtained through an iteratively reweighting least squares (IRLS), which is how it is implemented in \pkg{sprm}. In the package, the Fair, Huber or Hampel reweighting functions can be picked, which will lead to different robustness properties. For more detail on robust regression, the reader is referred to \cite{Hampel} or \cite{RousseeuwLeroy}. Note that a cellwise version of RM has recently been introduced \citep{CRM}.   
	
	\subsection{Sparse NIPALS}\label{sec:SNIPLS} 
	A second building block in the package is the SNIPLS algorithm. It is a sparse version of the NIPALS algorithm for PLS and as such, essentially a computationally efficient implementation of univariate sparse PLS \citep{ChunKeles2010}. Again, the SNIPLS components are linear combinations of the original variables through a set of weighting vectors $\mathbf{w}_i$ that maximize: 
	\begin{subequations}\label{eq:snipls}
		\begin{equation}\label{eq:snipls_ctricy}
		\mathbf{w}_i = \argmax_{\mathbf{a}} \mathop{\mbox{cov}^2}\left(\mathbf{a}^T\mathbf{X},\mathbf{y}\right) + \lambda \parallel\mathbf{a}\parallel_1, 
		\end{equation}
		subject to:
		\begin{equation}\label{eq:snipls_constrcy}
		\mathbf{w}_i^T\mathbf{X}^T\mathbf{X}\mathbf{w}_j = 0 \mbox{ and } \parallel \mathbf{w}_i\parallel_2 = 1, 
		\end{equation}
	\end{subequations}
	which in sparse PLS is typically maximized through a surrogate formulation. However, in this case, the exact solution to Criterion \eqref{eq:snipls} can be obtained, which is what the SNIPLS algorithm builds upon. For details on the algorithm, the reader is referred to \cite{SPRM-DA}. At this point, remark that the SNIPLS algorithm has also become a key building block to analyze outlyingness \citep{SPADIMO}. 
	
	\subsection{Sparse partial robust M}\label{sec:sprm-detail}
	Sparse partial robust M dimension reduction unites the benefits of SNIPLS and robust M estimation: is yields an efficient sparse PLS dimension reduction, while at the same time, it is robust against both leverage points and virtual outliers through robust M estimation. It is defined similarly as in \eqref{eq:snipls}, but instead maximizing a weighted covariance, with caseweights that depend on the data. Consistent with robust M estimation, it can be calculated through iteratively reweighting SNIPLS. The idea to iteratively reweight PLS goes back to \cite{CumminsAndrews1995}, but SPRM improves upon the original proposal by (i) yielding a sparse estimate, (ii) having a reweighting scheme as well as starting values that weight both in the score and residual spaces and (iii) by allowing different weight functions, the most tuneable one being the Hampel function. As mentioned in Section \ref{sec:ppreg}, all dimension reduction estimators can provide a basis to construct regularized regression, by performing regression of the dependent variable onto the space spanned by the variables estimated in the dimension reduction step. However, SPRM takes a different approach at this, where the reweighting scheme is wrapped around the SNIPLS procedure, and the regression step essentially takes place at the SNIPLS level. Note that at $\lambda = 0$, the method becomes non-sparse and reduces to PRM \citep{PRM}.
	
	\section{Cross-validation and plotting}\label{sec:cvplot}
	
	Each of the \pkg{sudire}, \pkg{ppdire} and \pkg{sprm} subpackages in \pkg{direpack} are wrappers around a broad class of dimension reduction methods. Each of these methods will have at least one tuneable hyperparameter; some have many more. The user will want to be able to find the optimal hyperparameters for the data at hand, which can be done through cross-validation or bayesian optimization. It is not the aim of \pkg{direpack} to provide its own hyperparameter tuning algorithms, as ample cross-validation utilities are available in \pkg{scikit-learn}'s \pkg{model\_selection} subpackage and the \pkg{direpack} estimations have been written consistently with the \pkg{scikit-learn} API, such that these model selection tools from \pkg{scikit-learn} can directly be applied to them. However, some caution should be taken when training the robust methods. While all classical (non-robust) methods could just use \pkg{scikit-learn}'s default settings, when tuning a robust model, outliers are expected to be in the data, such that it becomes preferable to apply a robust cross-validation metric as well. Thereunto, it is possible to use \pkg{scikit-learn}'s \texttt{median\_absolute\_error}, which is an MAE ($L_1$) scorer that is less affected by extreme values than the default \texttt{mean\_squared\_error}. However, particularly in the case of robust M estimators, a more model consistent approach can be pursued. The robust M estimators provide a set of case weights, and these can be used to construct a weighted evaluation metric for cross-validation. Exactly this is provided in the \texttt{robust\_loss} function that is a part of the \pkg{direpack} cross-validation utilities. 
	
	Similar to hyperparameter tuning, \pkg{direpack}'s mission is not to deliver a broad set of plotting utilities, but rather focus on the dimension reduction statistics. However, some plots many users would like to have in this context, are provided for each of the methods. These are 
	\begin{itemize}
		\item Projection plots. These plots visualize the scores $\mathbf{t}_i$ and a distinction can be made in the plots between cases that the model had been trained with, and test set cases. 
		\item Parity plots. For the regularized regressions based on the estimated scores, these visualize the predicted versus actual responses, with the same distinction as for the scores. 
	\end{itemize} 
	For the special case of SPRM, the plots have enhanced functionality. Since SPRM provides case weights, which can also be calculated for new cases, the SPRM plots can flag outliers. In the \texttt{sprm\_plot} function, this is set up with two cut-offs, based on the caseweight values, and visualized as {\em regular cases}, {\em moderate outliers} or {\em harsh outliers}. For SPRM, there is an option as well to visualize the caseweights themselves. 
	
	\begin{figure}
		\centering
		\includegraphics{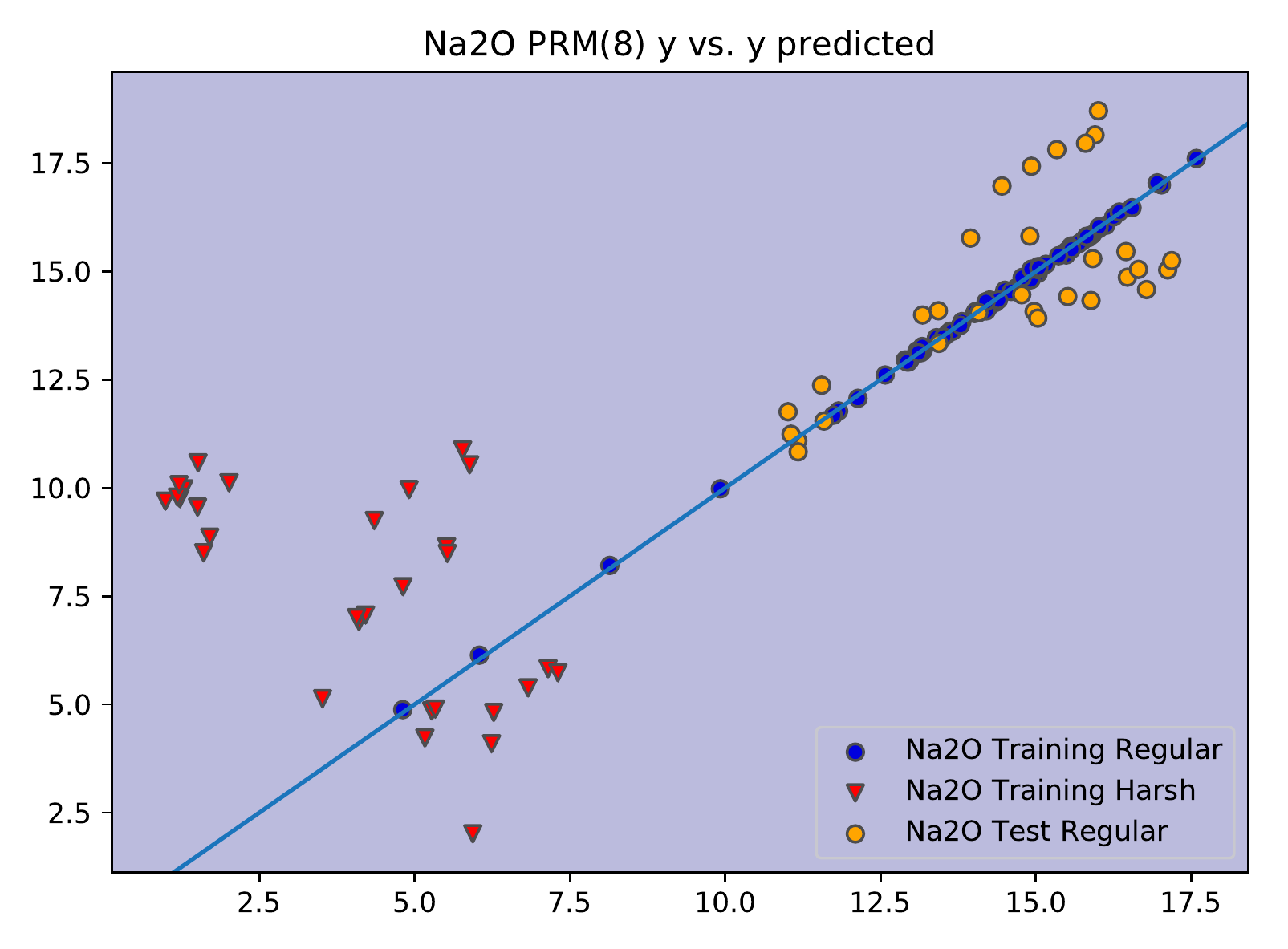}
		\caption{Parity plot for sodium oxide predictions in the glass data set, obtained from a PRM model with 8 latent variables, fit by \pkg{sprm} and plotted by \texttt{sprm\_plot}}
		\label{fig:glass_yyp}
	\end{figure}

	As an example, in Figure \ref{fig:glass_yyp} the parity plot is shown for one dependent variable (sodium oxide) for a specific subset of the {\em glass} data. These consist of EPXMA spectra and corresponding concentrations of a set of arch\ae ological glass vessels \citep{Janssens_Glass_1998}. These data have been investigated in several publications and it is not the focus of this article neither to describe the data in detail, nor how to analyze them. For this example, it suffices to know that the data are known to be heterogeneous and contain several clusters, as well as a set of outliers, some of which are related to the clusters and some not. To obtain the plot in Figure \ref{fig:glass_yyp}, cases 20 through 120 were taken as the training set and cases 121-150 were selected for the test set. It is noted that due to the structure of the data, much more diligent approaches could be pursued, but here it merely serves as an illustration of the plot functionality. A PRM model was fit to these data using \pkg{sprm} with eight latent variables, which has previously been reported to be the optimum number for the sodium oxide dependent variable. The parity plot, shown in \ref{fig:glass_yyp}, shows that the training cases fit the model well, except for some cases flagged as harsh outliers by PRM, such that these did not impact the model fit. These are plotted with a different symbol in \texttt{sprm\_plot}. The test cases, plotted in a different color, are all regular cases, yet it can be seen that there is more variance in model predictions for them, compared to the training set. 
	
	Finally, the authors would like to stress that this example has deliberately been kept to a minimal size. The package contains ample examples as \proglang{jupyter} notebooks, that can be consulted upon installing the package on its GitHub homepage, linked from PyPI.  
	
	\section{Conclusions and outlook} \label{sec:conclusions}
	In this article, \pkg{direpack} has been presented, a novel package that combines several classes of state-of-the-art dimension reduction techniques in one single package and format, consistent with the \pkg{scikit-learn} API for statistical and machine learning. The included classes of dimension reduction techniques are: projection pursuit dimension reduction, sufficient dimension reduction and robust M estimators for dimension reduction. As such, \pkg{direpack} is both the first package altogether that combines each of these classes of dimension reduction techniques in a single API, as well as the first package that brings these classes to the \proglang{Python} world. Moreover, \pkg{direpack} offers some novel pre-processing functions, as well as convenient tools for cross-validation and plotting.  
	
	Nothing of the like existed before \pkg{direpack} in \proglang{Python}, neither did it in other programming languages. While most of the methods in \pkg{direpack} are available in other languages, the reader might have to search thoroughly to find them, as some of the methods included in \pkg{direpack} are only available as \proglang{R} or \proglang{MATLAB} functions or packages solely distributed as supplementary materials to the corresponding publications, or through the author's personal or academic websites. Even for methods that are more widely available, without \pkg{direpack} it would still be a challenge to compare them in a benchmark study by setting up statistical learning pipelines. While the majority of the methods presented here, are also available as \proglang{R} packages on CRAN, the syntax in these packages varies widely and they cannot be included into \pkg{caret} pipelines without (sometimes major) modifications. By bringing all of these methods together in a single consistent framework, compatible with \pkg{scikit-learn}, the authors hope to promote advanced statistical dimension reduction learning in the \proglang{Python} community and bring these methods into more widespread use. 
	
	No package is complete and the authors would like to see \pkg{direpack} extend its functionality in the future. Some paths for further development could be: 
	\begin{itemize}
		\item {\em Cellwise robust estimation}. For instance, a cellwise robust version of the robust M regression method, included in \pkg{sprm}, has recently been published 
		\citep{CRM}, and could be included in \pkg{direpack}.
		\item {\em Uncertainty quantification}. The methods provided through \pkg{direpack} provide point estimates. In the future, the package could, e.g. be augmented with appropriate bootstrapping techniques, as was done for a related dimension reduction context \citep{SerneelsVanEspen2005}
		\item {\em GPU flexibility}. There are many matrix manipulations in \pkg{direpack}, which can possibly be sped up by allowing a GPU compatibility, which could be achieved by providing a \pkg{TensorFlow} or \pkg{PyTorch} back-end. However, this would be a major effort, since the present back-end integrally builds upon \pkg{numpy} (as does \pkg{scikit-learn}'s). 
	\end{itemize}
	
	While there always is ample room for development, the authors hope to have provided a novel package for dimension reduction data analytics in \proglang{Python} and hope that both \pkg{direpack} and the methods contained in it, will see more widespread adoption in the near future.

	\bibliography{bibS28}
	
	\appendix
	\section{Appendix: Jupyter Example Notebooks}\label{sec:app}
	The \pkg{direpack} package can be installed from its PyPI master source \url{https://pypi.org/project/direpack/} through \pkg{pip}: {\em \verb+pip install direpack+}
	\newline
	
	As mentioned before, the authors have kept the code snippets or illustrations to a minimal size, since the \pkg{direpack} package contains a set of \proglang{jupyter}  notebooks with examples for each category of methods:
	
	\begin{enumerate}
		\item {\em \verb+dicomo_example.ipynb+}: moment and co-moment estimators;
		\item {\em \verb+ppdire_example.ipynb+}: projection pursuit dimension reduction;
		\item {\em \verb+sprm_example.ipynb+}: robust M-estimators for dimension reduction;
		\item {\em \verb+sudire_example.ipynb+}: sufficient dimension reduction.
	\end{enumerate}
	
	In  {\em \verb+dicomo_example.ipynb+}, a toy example is loaded and then example code is provided to compute a wide range of product-moment statistics and energy statistics. 
	For variance, skewness and kurtosis estimation, a comparison with methods available in \pkg{numpy} and \pkg{scipy} is made. Co-moment estimation until fourth order is illustrated and some options for input parameters are described. For energy statistics, example code is given to estimate distance variance, distance covariance, martingale difference divergence, distance correlation, martingale difference correlation and distance continuum association. Where applicable, the results are compared to those obtained by the \pkg{dcor} package.
	
	In {\em \verb+ppdire.example.ipynb+}, different methods for projection pursuit dimension reduction using various optimization algorithms are illustrated on a toy example. For classical PCA and PLS, the  results thus obtained are compared to estimates obtained using \pkg{sklearn}. Example code to apply several robust projection pursuit estimators and to visualize the predicted versus the actual values and the obtained scores is given. Moreover, the computation of generalized betas, the application of cross-validation (through \pkg{scikit-learn}), as well as data compression, are illustrated in this notebook.
	
	The Sparse Partial Robust M-regression (SPRM estimator) is illustrated on real data in {\em \verb+sprm_example.ipynb+}. There is also separate example code for the building blocks of the SPRM estimator, namely the SNIPLS estimator and the robust M estimator. The plotting functionality is illustrated in detail and it is shown that all modules are compatible with \pkg{scikit-learn}'s APIs for model tuning. 
	
	The aim of {\em \verb+sudire_example.ipynb+} is to apply Sufficient Dimension Reduction techniques on real data. It is shown how a basis for the central subspace can be estimated via distance covariance, martingale difference divergence or a user defined function such as Ball covariance. Moreover, the estimation of the dimension of the central subspace is illustrated using directional regression, but other options are possible. The visualization tools are also demonstrated. 
	
\end{document}